\documentclass[twocolumn,pre,superscriptaddress, showpacs]{revtex4}
\usepackage{epsfig,graphicx,times}
\usepackage{amstext}
\usepackage{amsmath}
\usepackage{amssymb}
\usepackage{graphicx}
\usepackage{latexsym}
\usepackage{bm}
\usepackage{color}
\usepackage{subfigure}

\begin{document}

\title{Quantum Crooks fluctuation theorem and quantum Jarzynski equality in the presence of a reservoir}

\author{H. T. Quan}
\affiliation{Theoretical Division, MS B213, Los Alamos National Laboratory, Los Alamos,
NM, 87545, U.S.A.}
\author{H. Dong}
\affiliation{Institute of Theoretical Physics, Chinese Academy of
Sciences, Beijing, 100190, P.R. China}

\begin{abstract}
We consider the quantum mechanical generalization of Crooks
Fluctuation Theorem and Jarzynski Equality for an open quantum
system. The explicit expression for microscopic work for an
arbitrary prescribed protocol is obtained, and the relation between
quantum Crooks Fluctuation Theorem, quantum Jarzynski Equality and
their classical counterparts are clarified. Numerical simulations
based on a two-level toy model are used to demonstrate the validity
of the quantum version of the two theorems beyond linear response
theory regime.
\end{abstract}


\maketitle

\section{Introduction:}

Nonequilibrium thermodynamics has been an
intriguing research subject for more than one hundred years
\cite{nonequilibrium}. Yet our understanding about nonequilibrium
thermodynamic phenomena, especially about those far-from-equilibrium regime
(beyond the linear response regime), remains very limited. In the
past fifteen years, there are several significant breakthroughs in
this field, such as Evans-Searls Fluctuation Theorem \cite{evans},
Jarzynski Equality (JE) \cite{JE}, and Crooks Fluctuation Theorem
(Crooks FT) \cite{crooks}. These new theorems not only have
important applications in nanotechnology and biophysics, such as
extracting equilibrium information from nonequilibrium measurements,
but also shed new light on some fundamental problems, such as improving our understanding of how the thermodynamic
reversibility arise from the underlying time reversible dynamics.

Since the seminal work by Jarzynski and Crooks a dozen of years ago,
the studies of nonequilibrium thermodynamics in small system
attract numerous attention \cite{followup}, and the validity and
universality of these two theorems in classical systems has been extensively studied not only by numerical studies \cite{classicalnumerical},
but also by experimental exploration \cite{classicalexperiment} in
single RNA molecules, and for both deterministic and stochastic processes. For quantum systems, possible quantum
extension of Crooks FT and JE have also been reported
\cite{quantumJE}. Nevertheless, we notice that almost all of these reports about
quantum extension of Crooks FT focus on isolated quantum systems
\cite{isolatedcrooks}, and the explicit expression of microscopic
work, and their distributions in the presence of a heat bath are not extensively studied. In addition,
the relationship between classical and quantum Crooks FT is not
addressed adequately so far. As a result, the experimental studies of quantum Crooks FT
and JE are not explored (an exception is the experimental scheme of
quantum JE of isolated system based on trapped ions \cite{schmidt}).

%Figure 1
%\begin{figure}[h]
\begin{figure}[ht]
\begin{center}
\includegraphics[width=8cm, clip]{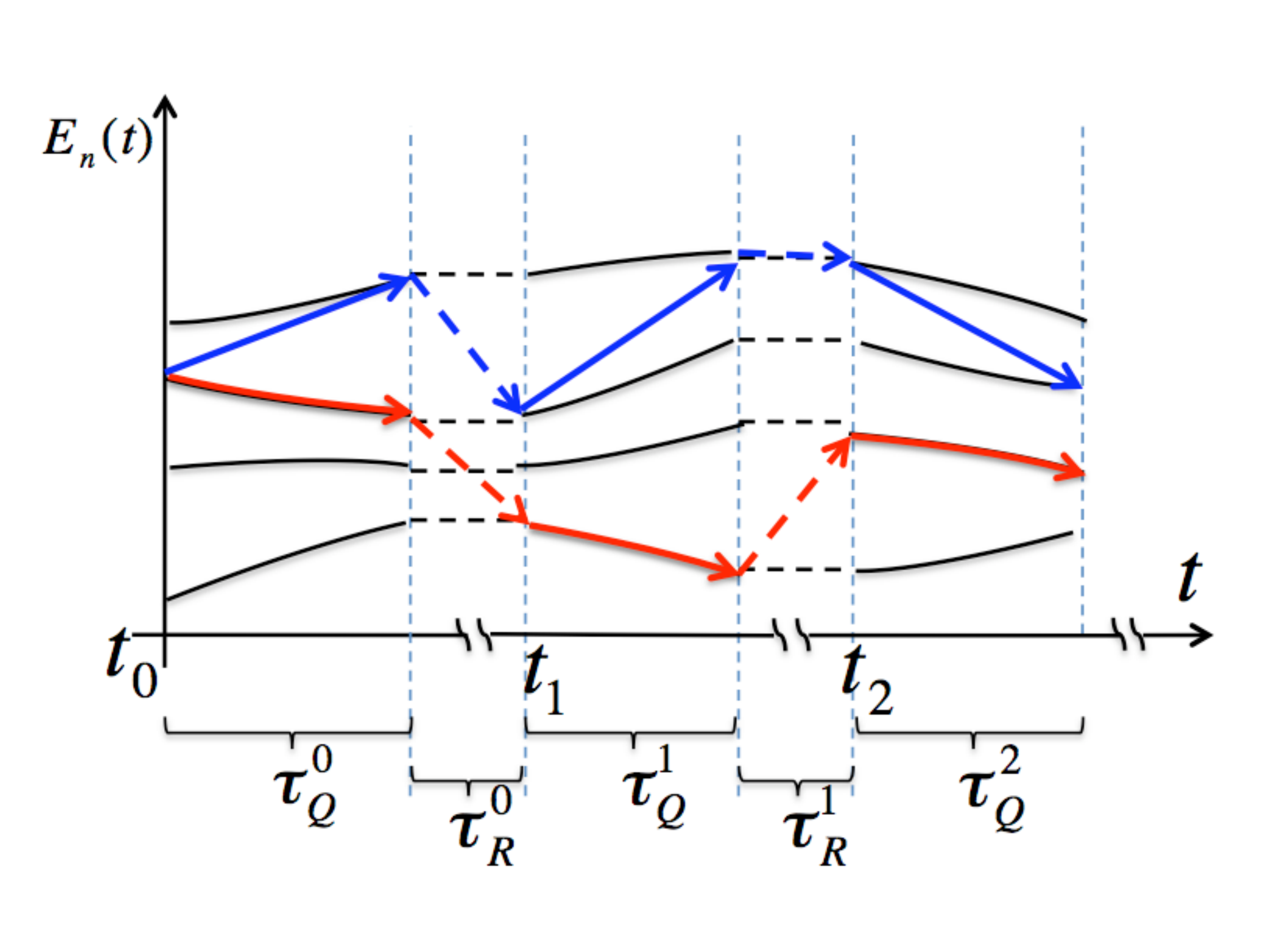}
\end{center}
\caption{(\textit{Color Online}) Trajectories of a quantum system in
a nonequilibrium process. Similar to Ref. \cite{crooks} each step
(from $t_{n}$ to $t_{n+1}$) is divided into two substeps: the
controlling substep of time $\tau_{Q}^{n}$, in which the energy
spectrum (black solid line) of the system change with time, and the
relaxation substep of time $\tau_{R}^{n}$ in which the energy
spectrum (black dashed line) remains unchanged. In the controlling
substep (solid line) work is done, but there is no heat exchange;
While in the relaxation substep, there is heat exchange between the
system and the heat bath, but there is no work done. Blue trajectory
corresponds to fast controlling protocol, during which there are usually
interstate excitations in the controlling substep. Red trajectory corresponds to slow (quantum
adiabatic) controlling protocol, and the system remains in its
instantaneous eigenstate in the controlling substep. Red trajectory
is the counterpart of classical case.} \label{fig1}
\end{figure}

In this paper, we will give a detailed proof of the validity of
quantum Crooks FT and quantum JE for an open quantum system based on
the explicit expression of microscopic work and their corresponding
probability distributions for an arbitrary prescribed controlling protocol. We also clarify the relation between
quantum Crooks FT, quantum JE and their classical counterparts. In
the last part of the paper, the studies based on a two-level system
are given as an illustration to demonstrate our central idea.

\section{Notations and assumptions:}

Crooks FT \cite{crooks} is firstly derived in classical systems in a
microscopically reversible Markovian stochastic process. In the
proof of a classical Crooks FT, a key technique is to separate work
steps from heat steps. In the following discussion of quantum
extension of Crooks FT and JE, we will employ the same technique as
that used in Ref. \cite{crooks} to separate the controlling process
into two substeps: controlling substep and relaxation substep (see
Fig. 1). The controlling substep proceeds so quickly in comparison
with the thermalization process of the system that we can ignore the
influence of the heat bath during the controlling substep. So there
is only work done in the controlling substep. In the relaxation
substep, on the other hand, there is only heat exchange.

Having clarified the main strategy (separating work substep from
heat substep), let us come to the details of the notations and
assumptions. We employ the same notations and assumptions as that in
Ref. \cite {crooks} to prove the quantum Crooks FT. In Ref.
\cite{crooks} the author assumes discrete time and discrete phase
space. Here, the discrete energy spectrum in a quantum system in
place of the discrete phase space of a classical system arises
naturally. We also assume discrete time $t_{0}$, $t_{1}$, $t_{2}$,
$t_{3}$, $\cdots$, $t_{N}$ for the quantum system (see Fig. 1). The
parameter $\lambda(t)$ is controlled according to an arbitrary
prescribed protocol $\lambda (t_{0})=\lambda_{A}$, $\lambda
(t_{1})=\lambda_{1}$, $\lambda (t_{2})=\lambda_{2}$, $\cdots$,
$\lambda (t_{N})=\lambda_{B}$, where $A$ and $B$ depict the initial
and final points of the process. Every step $t_{n}\rightarrow
t_{n+1}$ is seperated into controlling substep of time $\tau_{Q}^{i}$
and relaxation substep of time time $\tau_{R}^{i}$,
$t_{i+1}=t_{i}+\tau_{Q}^{i}+\tau_{R}^{i}$ (see Fig. 1). If we use
$\left \vert i_{n}, \lambda_{m} \right \rangle$ and $E(i_{n},
\lambda_{m})$ to depict the $i_{n}$-th instantaneous eigenstate and
eigenenergy of the system Hamiltonian $H(\lambda_{m})$, we can
rewrite the trajectory $A\rightarrow B$ of Ref. \cite{crooks} in the
following way
\begin{equation}
\begin{split}
\left \vert i_{0},\lambda_{0} \right \rangle \rightarrow \left \vert
i_{0},\lambda_{1}\right \rangle \underrightarrow{\lambda_{1}} \left \vert
i_{1},\lambda_{1}\right \rangle \rightarrow \left \vert
i_{1},\lambda_{2}\right \rangle \underrightarrow{\lambda_{2}}
\left \vert i_{2},\lambda_{2}\right \rangle \\
\rightarrow \cdots \rightarrow \left \vert i_{N-1},\lambda_{N-1}\right
\rangle \rightarrow \left \vert i_{N-1},\lambda_{N}\right \rangle
\underrightarrow{\lambda_{N}} \left \vert i_{N},\lambda_{N}\right\rangle.
\end{split}
\end{equation}
In the classical case, the system remains in its $i_{n}$-th state of
the discrete phase space during the controlling substep. Analogously, in
quantum systems, this process corresponds to the quantum adiabatic
regime, i.e., the system remains in its $i_{n}$-th eigenstate of
the instantaneous Hamiltonian when we control the parameter
$\lambda(t)$ of the Hamiltonian $H[\lambda(t)]$ so slowly that the
quantum adiabatic conditions are satisfied, and the above
trajectories (1) can be achieved (red trajectory of Fig. 1). However, if we control the
parameter of the Hamiltonian very quickly in the controlling substep, and then the quantum
adiabatic conditions are not satisfied, the trajectory $A\rightarrow B$ in general
should be written as (see blue trajectory of Fig. 1)
\begin{equation}
\begin{split}
\left \vert i_{0},\lambda_{0} \right \rangle \rightarrow \left \vert
i_{0}^{\prime},\lambda_{1}\right \rangle \underrightarrow{\lambda_{1}} \left \vert
i_{1},\lambda_{1}\right \rangle \rightarrow \left \vert
i_{1}^{\prime},\lambda_{2}\right \rangle \underrightarrow{\lambda_{2}}
\left \vert i_{2},\lambda_{2}\right \rangle \\
\rightarrow \cdots \rightarrow \left \vert i_{N-1},\lambda_{N}\right
\rangle \rightarrow \left \vert i_{N-1}^{\prime},\lambda_{N}\right
\rangle \underrightarrow{\lambda_{N}} \left \vert i_{N},\lambda_{N}\right\rangle.
\end{split}
\end{equation}
The main difference of the above two kinds of trajectories (1) and
(2) is that after the controlling substep the system may not be in
the same eigenstate as that before the controlling, i.e., $i_{n}\neq
i^{\prime}_{n}$. The internal excitation $\left \vert
i_{n},\lambda_{n} \right \rangle \rightarrow \left \vert
i_{n}^{\prime},\lambda_{n+1}\right \rangle $ is due to randomness
caused by quantum non-adiabatic transition and has no classical
counterpart. Actually this difference of trajectories (1) and (2)
highlights the main difference between the quantum and classical
Crooks FT. For a quantum system, the microscopic work done in every
controlling substep is equal to the difference of the energy before
and after the controlling substep: $W_{n}=E(i_{n}^{\prime},
\lambda_{n+1})-E(i_{n}, \lambda_{n})$, and the heat exchanged with
the heat bath is equal to the difference of the energy of the system
before and after the relaxation substep $Q_{n}=E(i_{n},
\lambda_{n})-E(i_{n-1}^{\prime},\lambda_{n})$. For the trajectory
(2) as a whole, we must make $2N$ times quantum measurements to
confirm the microscopic work done and heat exchanged with the heat
bath. Similar to the classical case, the total work $W$ performed on
the system, and the total heat $Q$ exchanged with the heat bath are
given by the summation of work and heat in every step,
$W=\sum_{n=0}^{N-1} \left[E(i_{n}^{\prime}, \lambda_{n+1})-E(i_{n},
\lambda_{n})\right], Q=\sum_{n=0}^{N}\left[E(i_{n},
\lambda_{n})-E(i_{n-1}^{\prime}, \lambda_{n})\right]$, and the total
change in energy is $\Delta E=Q+W=E(i_{N}, \lambda_{N})-E(i_{0},
\lambda_{0})$. Note that the work and heat depend on the trajectory,
but the energy change depends only on the initial and final energy,
and does not depend on the trajectory.

Similar to the classical case \cite{crooks} we assume the trajectory
(2) to be Markovian, and the forward process starts from the thermal
equilibrium distribution $P( \left\vert i_{0}, \lambda_{0} \right
\rangle )=e^{-\beta E(i_{0}, \lambda_{0})}/(\sum_{i} e^{-\beta E(i,
\lambda_{0})})$. The joint probability for a given trajectory (2)
can be expressed as
\begin{equation}
\begin{split}
P_{F}(A \rightarrow B)=& P(
\left \vert i_{0},\lambda_{0}\right \rangle) \prod_{n=0}^{N-1} P_{F}(
\left \vert i_{n},\lambda_{n}\right \rangle \rightarrow \left \vert
i_{n}^{\prime},\lambda_{n+1}\right \rangle)\\
& \times P_{F}( \left \vert i_{n}^{\prime},\lambda_{n+1}\right
\rangle \rightarrow \left \vert i_{n+1},\lambda_{n+1}\right
\rangle).  \label{3}
\end{split}
\end{equation}
It can be seen that the above probability (3) of a trajectory for a
quantum case is different from the classical case \cite{crooks} by
the extra term $P( \left \vert i_{n},\lambda_{n}\right \rangle
\rightarrow \left \vert i_{n}^{\prime},\lambda_{n+1}\right \rangle)$
arising from randomness due to quantum non-adiabatic transition.
When the quantum adiabatic conditions are satisfied, $P( \left \vert
i_{n},\lambda_{n}\right \rangle \rightarrow \left \vert
i_{n}^{\prime},\lambda_{n+1}\right \rangle)= \delta_{i_{n},i_{n}^{\prime}}$, we regain the
probability of a trajectory in classical systems \cite{crooks}. We
will see later that the quantum Crooks FT and quantum JE in the
quantum adiabatic regime are the counterpart of classical Crooks FT
and classical JE.

To prove the quantum Crooks FT, we also need to consider the
time-reversed trajectory \cite{reverse} of the original trajectory
(2). The time-reversed trajectory corresponding to the forward time
trajectory $A \leftarrow B$ in Eq. (2) can be written as
\begin{equation}
\begin{split}
\Theta \left \vert i_{0},\lambda_{0} \right \rangle \leftarrow
\Theta\left \vert i_{0}^{\prime},\lambda_{1}\right \rangle
\underleftarrow{\lambda_{1}} \Theta\left \vert i_{1},\lambda_{1}\right \rangle
\leftarrow
\Theta\left \vert i_{1}^{\prime},\lambda_{2}\right \rangle \underleftarrow{\lambda_{2}} \\
 \cdots \leftarrow
\Theta\left \vert i_{N-1},\lambda_{N}\right \rangle \leftarrow
\Theta\left \vert i_{N-1}^{\prime},\lambda_{N}\right \rangle
\underleftarrow{\lambda_{N}} \Theta\left \vert i_{N},\lambda_{N}\right\rangle
\end{split}
\end{equation}
where $\Theta \left \vert i_{n},\lambda_{n} \right \rangle = \left
\vert i_{n},\lambda_{n} \right \rangle^{\ast} $ is the microscopic
state in the time-reversed trajectory \cite{sakurai}. The sequence
in which states are visited is reversed, as is the order in which
$\lambda$ is changed. The work done $W$, the heat exchange $Q$ with
the heat bath, the change of the internal energy $\Delta E$, and the
change of free energy $\Delta F$ for the reversed time direction are
the negative value of that of the forward time trajectory. The joint
probability for time reversed trajectory $A \leftarrow B$ can be expressed as
\begin{equation}
\begin{split}
P_{R}(A \leftarrow B)= &\prod _{n=0}^{N-1} P_{R}(\Theta \left \vert
i_{n},\lambda_{n} \right \rangle \leftarrow \Theta\left \vert
i_{n}^{\prime},\lambda_{n+1}\right \rangle) \\
&\times  P_{R}(\Theta\left \vert i_{n}^{\prime},\lambda_{n+1}\right \rangle \leftarrow \Theta\left
\vert i_{n+1},\lambda_{n+1}\right \rangle) \\
&\times P(\Theta\left \vert i_{N},\lambda_{N}\right \rangle),
\end{split}
\end{equation}
where $P( \Theta \left\vert i_{N}, \lambda_{N} \right \rangle
)=e^{-\beta E(i_{N}, \lambda_{N})}/\sum_{i} e^{-\beta E(i,
\lambda_{N})})$ is the initial thermal distribution for the
time-reversed trajectory. Also there is en extra term $P_{R}(\Theta
\left \vert i_{n},\lambda_{n} \right \rangle \leftarrow \Theta\left
\vert i_{n}^{\prime},\lambda_{n+1}\right \rangle)$ arising due to
the randomness caused by quantum non-adiabatic transition in
comparison with the classical case.

\section{Proof of quantum crooks FT and quantum JE}

As we have mentioned
before, in a trajectory every step consists of two substeps, the controlling substep
(not necessarily to be quantum adiabatic) and the relaxation
substep. The relaxation substeps are assumed to be microscopically
reversible, and therefore obey the detailed balance \cite
{crooks,chandler} for all fixed value of the external control
parameter $\lambda$
\begin{equation}
\frac{P_{F}( \left\vert i_{n-1}^{\prime}, \lambda_{n} \right \rangle
\rightarrow \left \vert i_{n}, \lambda_{n}\right \rangle
)}{P_{R}(\Theta\left \vert i_{n-1}^{\prime}, \lambda_{n} \right
\rangle \leftarrow \Theta\left \vert i_{n},
\lambda_{n}\right\rangle)}=\frac{e^{-\beta
E(i_{n},\lambda_{n})}}{e^{-\beta E(i_{n-1}^{\prime},\lambda_{n})}}.
\end{equation}
To compare the ratio of the probabilities of forward (3) and
time-reversed (5) trajectories, we also need to know the ratio of
the probabilities in the controlling substep. In the following we
will focus on the study of controlling substep and its time
reversal. As we mentioned before, during the controlling substep,
the system can be regarded as an isolated quantum system and the
evolution is completely determined by a time-dependent Hamiltonian
$H[\lambda(t)]$. For example, when the controlling parameter
$\lambda$ is changed from $\lambda_{n}$ to $\lambda_{n+1}$, the
probability of the transition from a microscopic state $\left \vert
i_{n},\lambda_{n}\right\rangle$ to another microscopic state $\left
\vert i_{n}^{\prime},\lambda_{n+1}\right\rangle$ can be expressed as
\begin{equation}
P_{F}( \left\vert i_{n},\lambda_{n}\right\rangle \rightarrow
\left\vert i_{n}^{\prime},\lambda_{n+1}\right\rangle)=|\left\langle
i_{n}^{\prime},\lambda_{n+1}\right\vert U \left \vert
i_{n},\lambda_{n} \right \rangle |^2
\end{equation}
where $U=\mathrm {T} \exp\{- i \int_{t_{0}}^{t_{1}}
H[\lambda(t)]dt\}$ is the unitary matrix describing the evolution of
the isolated quantum system in the controlling substep, and $\mathrm
{T}$ is the time-ordered operator. Similarly, in the time-reversed
trajectory the excitation probability from the microscopic state
$\Theta \left\vert i_{n}^{\prime},\lambda_{n+1}\right\rangle$ to
another microscopic state $\Theta \left\vert i_{n},\lambda_{n}
\right\rangle$ in the time reversed trajectory can be expressed as
\cite{theta}
\begin{equation}
\begin{split}
P_{R}(\Theta & \left\vert i_{n},\lambda_{n} \right\rangle \leftarrow
\Theta \left\vert  i_{n}^{\prime},\lambda_{n+1}\right\rangle )\\
& =| \left ( \left\langle i_{n},\lambda_{n}\right\vert
\overleftarrow{\Theta} \right ) \Theta U\overleftarrow{\Theta}
\left (\Theta \left \vert i_{n}^{\prime},\lambda_{n+1} \right
\rangle \right) |^2,
\end{split}
\end{equation}
where $\Theta U\overleftarrow{\Theta}=\mathrm{T} \exp\{-i
\int_{t_{0}}^{t_{1}} H[\lambda(t_{0}+t_{1}-t)]dt\}=
(U^{\dagger})^{\ast}= U^{T}$ is the time-reversed unitary matrix.
Because of the property of the time-reversed transformation $\Theta
\left\vert i_{n},\lambda_{n} \right\rangle=\left\vert
i_{n},\lambda_{n} \right\rangle^{\ast}$, and the property of the
Hermitian conjugate matrix,
\begin{equation}
(\left\langle i_{n},\lambda_{n}\right\vert)^{\ast} U^{T} (\left \vert
i_{n}^{\prime},\lambda_{n+1} \right \rangle )^{\ast} \equiv \left\langle
i_{n}^{\prime},\lambda_{n+1} \right\vert U \left \vert
i_{n},\lambda_{n} \right \rangle
\end{equation}
it is not difficult to prove that
\begin{equation}
\frac{P_{F}(\left\vert i_{n},\lambda_{n} \right \rangle\rightarrow
\left\vert i_{n}^{\prime},\lambda_{n+1}\right \rangle)}{P_{R}(\Theta
\left\vert i_{n},\lambda_{n} \right \rangle \leftarrow \Theta
\left\vert i_{n}^{\prime},\lambda_{n+1}\right \rangle )}\equiv 1.
\end{equation}
Based on the above two results (6), (10) and Eqs. (3) and (5), we reproduce the Crooks
FT for a quantum mechanical system
\begin{equation}
\frac{P_{F}(A \rightarrow B)}{P_{R}(A \leftarrow
B)}=e^{\beta (W- \Delta F)}.
\end{equation}
From Eq. (11) we group all those trajectories with the same amount of microscopic work, and obtain
\begin{equation}
\frac{P_{F}(W|_{a})}{P_{R}(-W|_{-a})}=e^{\beta (a- \Delta F)}.
\end{equation}
Eq. (12) is the Crooks FT. Similar to the derivation in Ref. \cite{crooks}, we obtain the JE
for a quantum open system straightforwardly $\left\langle e^{-\beta W}
\right\rangle = e^{-\beta \Delta F}$ from $\int P_{R}(-W|_{-a}) d a=1 $. Here, we would like to
emphasize that though quantum generalization of Crooks FT and JE
have been reported in some previous work, the explicit consideration
of the influence of the heat bath, i.e., the explicit expression of
microscopic work in the presence of a heat bath has not been
reported before. Also the relation between quantum and classical
trajectories are not addressed clearly. Hence our quantum mechanical
extensions of Crooks FT and JE are highly nontrivial.

\section{Illustration of quantum Crooks FT and quantum JE in a two-level
system}

%Figure 2
%\begin{figure}[h]
\begin{figure}[ht]
\begin{center}
\includegraphics[width=8cm, clip]{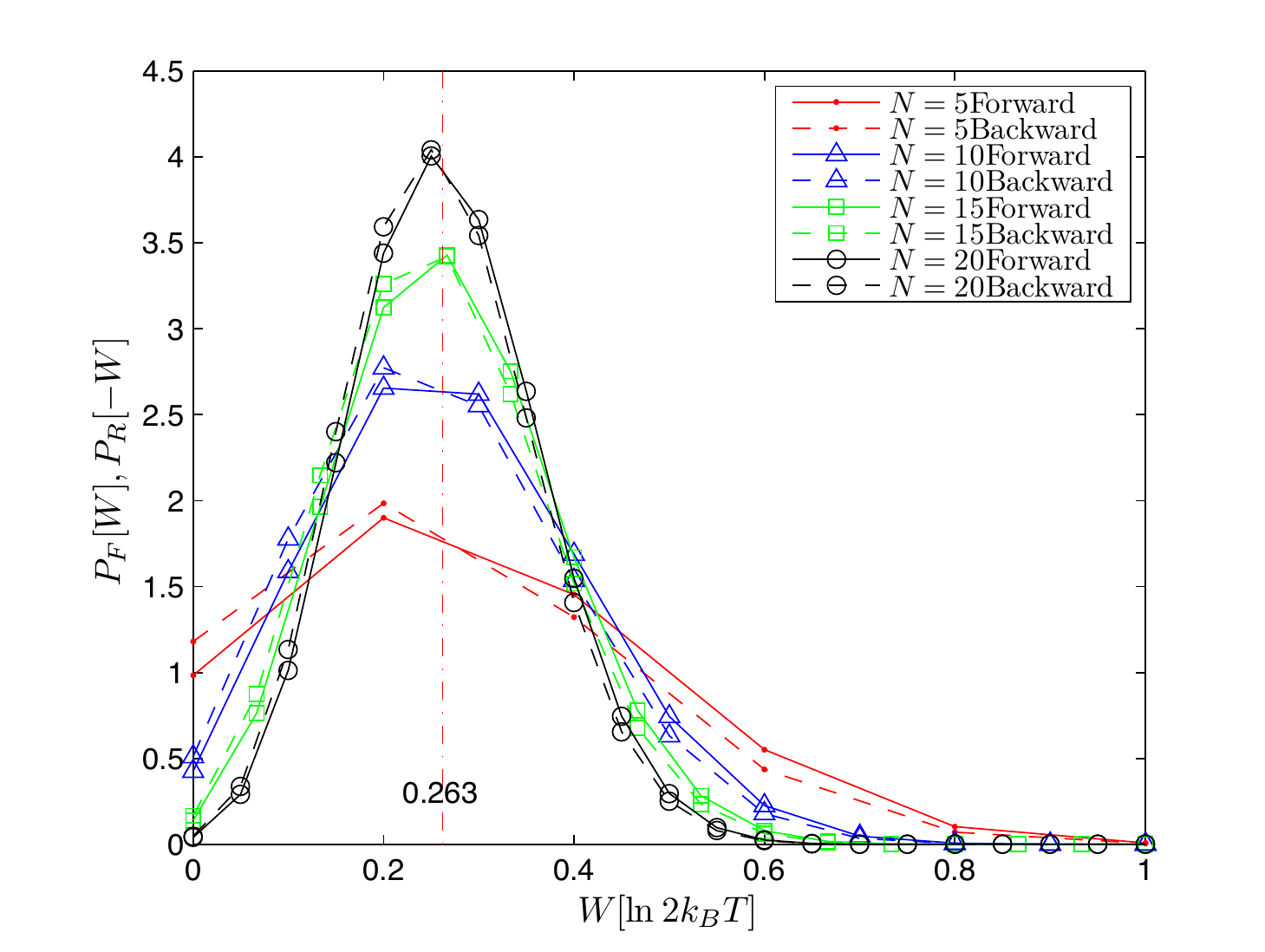}
\end{center}
\caption{(\textit{Color Online}) Microscopic work distribution
$P_{F}(W)$ of forward trajectories
 (solid lines), and the negative reverse work distribution $P_{R}(-W)$ of  their corresponding time-reversed
trajectories (dashed lines). The probabilities have been normalized.
Here we fix $\Delta(t_{0})$ and $\Delta(t_{N})$. Different
distributions represent different controlling time (the more steps,
the longer control time). The controlling steps are chosen to be
$N=5$ (red $\bullet$),  $N=10$ (blue $\bigtriangleup$), $N=15$
(green $\square$), and $N=20$ (black $\bigcirc$). It can be seen
that the work distributions for both forward and reversed
trajectories are not Gaussian. Moreover, with the decrease of the
controlling speed, the fluctuation of the distributions decreases,
and the difference between the work distribution of the forward
and time-reversed trajectories becomes less obvious. The
corresponding forward and negative reverse work distribution cross
at $W= \Delta F$, and this is a direct consequence of the quantum
Crooks FT. The free energy difference $\Delta F$ ia marked by the
red vertical dash-dotted line. } \label{fig2}
\end{figure}

Having generalized the Crooks FT and JE to quantum systems in the
presence of a heat bath. In the following, we use the studies based
on a two-level system \cite{quan08} as an illustration to
demonstrate our main idea. The Hamiltonian of the two-level system
is $H=\Delta(t)\left(\sigma_{z}+1\right)/2$, where $\Delta(t)$ is
the parameter of the Hamiltonian, and $\sigma_{z}$ is Pauli matrix.
The initial and final value of the parameter are
$\Delta_{A}=\Delta(t_{0})$ and $\Delta_{B}=\Delta(t_{N})$
respectively. The controlling scheme is the same as that in Ref.
\cite{quan08}: We divide the whole process into N even steps. Hence
the parameter in the $n$th step is
$\Delta(t_{n})=\Delta(t_{0})+n\Delta$, $n=1$, $2$, $\cdots$, $N$, where
$\Delta=(\Delta_{B}-\Delta_{A})/N$ is the change of the parameter in
every step. Every step consists of two substeps: the controlling
substep, in which we change the parameter from $\Delta(t_{n})$ to
$\Delta_{n+1}=\Delta(t_{n})+\Delta$, and the relaxation substep. For
simplicity, we consider the case where the system reaches thermal
equilibrium with the heat bath in every relaxation substep. Hence,
the probability for the forward and reverse relaxation substep can
be expressed as $P_{F}( \left\vert i_{n-1}^{\prime}, \lambda_{n}
\right \rangle \rightarrow \left \vert i_{n}, \lambda_{n}\right
\rangle )=e^{-\beta E(i_{n}, \lambda_{n})}/(\sum_{i} e^{-\beta E(i,
\lambda_{n})})$, and $P_{R}(\Theta\left \vert i_{n-1}^{\prime},
\lambda_{n} \right \rangle \leftarrow \Theta\left \vert i_{n},
\lambda_{n}\right\rangle)=e^{-\beta E(i_{n-1}^{\prime},
\lambda_{n})}/(\sum_{i} e^{-\beta E(i, \lambda_{n})})$. Also we
assume the quantum adiabatic conditions are satisfied in every
controlling substep. That is $P_{F}(\left\vert i_{n},\lambda_{n}
\right \rangle\rightarrow \left\vert
i_{n}^{\prime},\lambda_{n+1}\right \rangle)= \delta_{i_{n},i_{n}^{\prime}}$, and $P_{R}(\Theta
\left\vert i_{n},\lambda_{n} \right \rangle \leftarrow \Theta
\left\vert i_{n}^{\prime},\lambda_{n+1}\right \rangle )= \delta_{i_{n},i_{n}^{\prime}}$. Based on
these assumptions, the microscopic work distribution for the forward
trajectories can be obtained \cite{quan08}
\begin{equation}
P_{F}(W|_{k\Delta})=P^{F}_{e} \prod_{l=0}^{N-k-1}\frac{e^{\beta \Delta
_{B}}-e^{\beta (\Delta _{A}+l\Delta )}}{e^{\beta (l+1)\Delta }-1},
\end{equation}
where
\begin{equation}
P^{F}_{e}= \prod_{j=1}^{N}  \frac{e^{-\beta [\Delta _{A}
+(j-1)\Delta ]} }{1+e^{-\beta [\Delta _{A} +(j-1)\Delta ]}}, k=0, 1,
2, \cdots, N.
\end{equation} Similarly, the microscopic
work distribution for the time-reversed trajectory can be expressed
as
\begin{equation}
P_{R}(-W|_{-k\Delta})=P^{R}_{e} \prod_{l=0}^{N-k-1}\frac{e^{\beta \Delta}[e^{\beta \Delta
_{B}}-e^{\beta (\Delta _{A}+l\Delta )}]}{e^{\beta (l+1)\Delta }-1},
\end{equation}
where
\begin{equation}
P^{R}_{e} = \prod_{j=1}^{N}  \frac{e^{-\beta [\Delta _{B}
-(j-1)\Delta ]} }{1+e^{-\beta [\Delta _{B} -(j-1)\Delta ]} }, k=0,
1, 2, \cdots, N.
\end{equation}  We plot the above distributions (13)
and (14) of microscopic work in Fig. 2. Here the probability
distribution in the excited state are $P_{e}(\Delta_{A}) =e^{-\beta
\Delta_{A}}/(1+ e^{-\beta \Delta_{A}}) = 1/3$, and
$P_{e}(\Delta_{B})=e^{-\beta \Delta_{B}}/(1+ e^{-\beta \Delta_{B}})=
1/5$. The free energy difference is $\Delta F_{AB}=\left[
\ln(1+1/2)- \ln(1+1/4) \right] k_{B}T \approx 0.263\ln2 k_{B}T $. It
can be seen (see Fig. 2) that the corresponding forward and negative
reverse work distributions cross at $W= \Delta F$, no matter what
the controlling protocol is, and this result is a direct consequence
of Crooks FT. It should be pointed out that the work
distributions (13) and (15) are non-Gaussian \cite{quan08}. Hence, the processes
discussed here are beyond the linear response regime. Yet we will see
both Crooks FT and JE holds. We also plot the logarithm of the ratio
of the forward and negative reverse work distribution (See Fig.
3(a)). It can be seen that all data collapse onto the same straight
line. In addition, the slope of the line is equal to unit,
 and the line cross the horizontal axis at $W=0.263\ln2 k_{B}T=\Delta F_{AB}$. Thus
our numerical simulation confirms the validity of quantum Crooks FT
when the process is beyond the linear response regime. We also plot
the logarithm of the exponent averaged work $\ln \left\langle
e^{-\beta W}\right \rangle$ and averaged work $\left\langle W
\right\rangle$ of the forward process (see Fig. 3(b)) to test the
validity of quantum JE. It can be seen that the averaged work is
greater than the free energy difference $\left\langle W
\right\rangle \geqslant \Delta F$, while the logarithm of the
exponent averaged work is identical to the difference of the free
energy $\ln \left\langle e^{-\beta W}\right\rangle \equiv \Delta
F\approx 0.1823 k_{B}T$ no matter what the controlling protocol is.
Hence, Fig. 3(b) verifies quantum JE when the process is beyond the
linear response regime.

%Figure 3
%\begin{figure}[h]
\begin{figure}[ht]
\begin{center}
\subfigure[]{\includegraphics[width=8cm, clip]{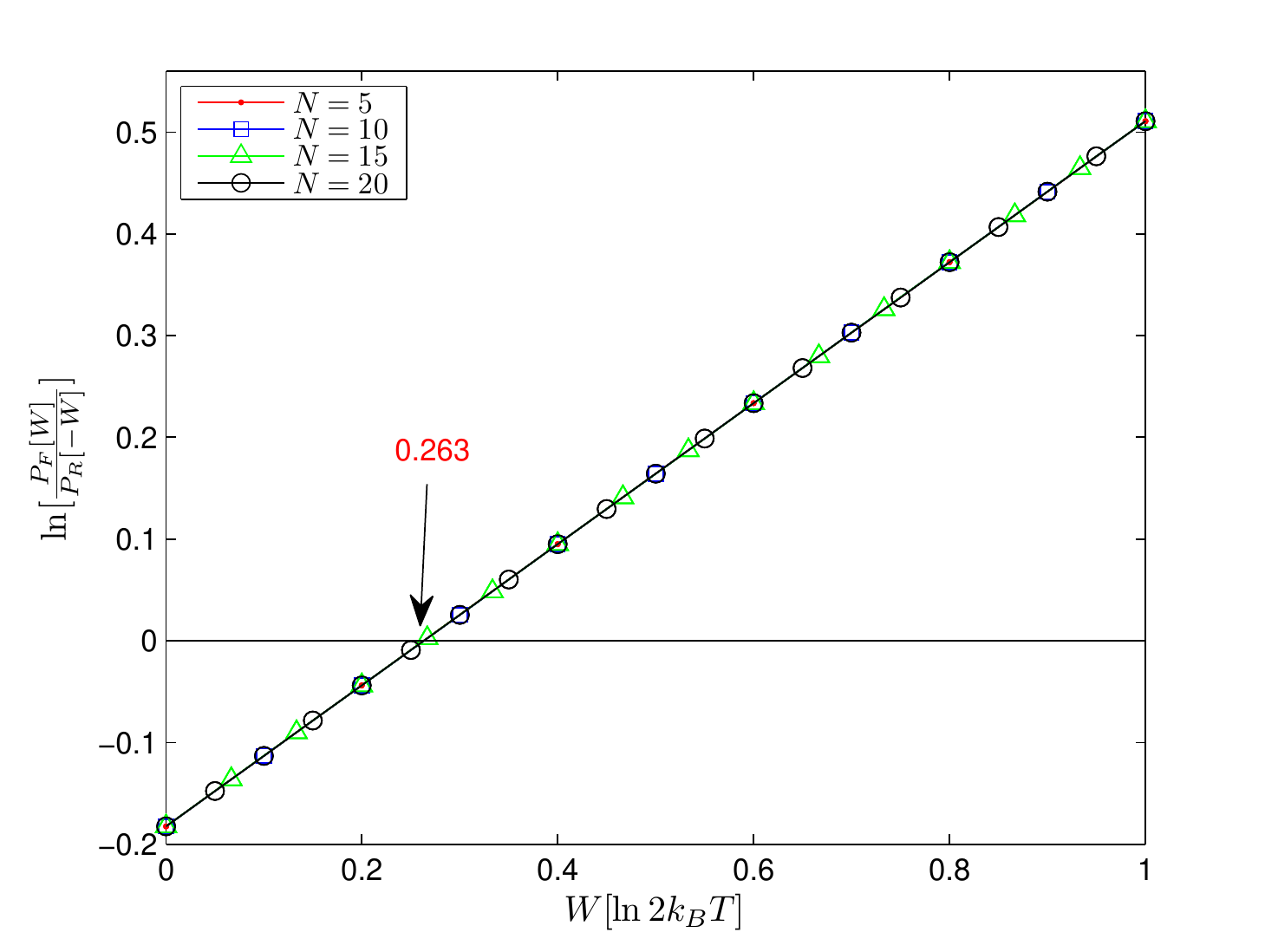}}
\subfigure[]{\includegraphics[width=8cm, clip]{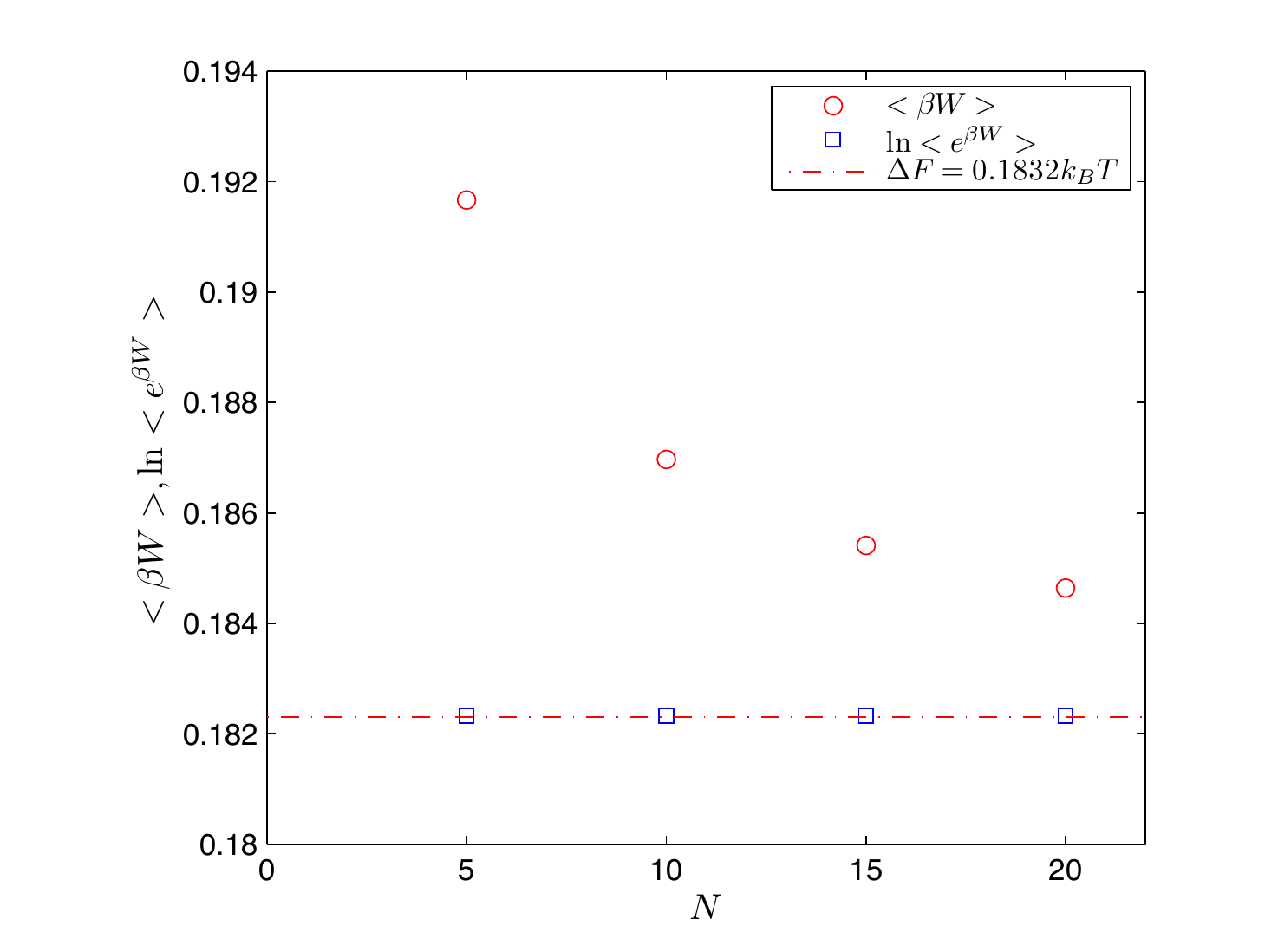}}
\end{center}
\caption{(\textit{Color Online}) (a) The logarithm of the
probabilities of forward and time-reversed trajectories as a
function of work. It can be seen that all data of different work and
different control protocols ($N=5$ (red $\bullet$),  $N=10$ (blue
$\bigtriangleup$), $N=15$ (green $\square$), and $N=20$ (black
$\bigcirc$) ) collapse onto the same straight line. The slop of the
line is equal to unity, and the line cross the horizontal axes at
$W= \Delta F$. Thus the numerical result verifies the quantum Crooks
FT $\ln \left[ P_{F}(W|_{a})/P_{R}(-W|_{-a})\right]=\beta(a- \Delta
F)$. (b) The averaged work VS. the logarithm of averaged exponent
work for different control protocols. It can be seen that the
averaged work $\left\langle W \right\rangle$ (red $\bigcirc$) is
always greater than the difference of free energy $\Delta F_{AB}$
and differ from one control protocol to another, while the logarithm
of the exponentially averaged work $\ln \left\langle \exp[-\beta
W]\right\rangle$ (blue $\square$) is always equivalent to the
difference of free energy irrespective of the control protocols.
Thus the numerical result verifies the JE $\ln{\left \langle
\exp[-\beta W]\right \rangle} \equiv \Delta F$.} \label{fig3}
\end{figure}

\section{Conclusion and remarks}

In this paper, we explicitly consider the quantum Crooks FT and
quantum JE in the presence of an external heat bath. Our proof
includes the proof of classical Crooks FT as a special case. When
the quantum adiabatic conditions are satisfied, we reproduce the
result of Crooks FT and JE for classical systems. Our work indicates
that in quantum systems, the probabilities (Eqs. (3) and (5)) comes
from the quantum non-adiabatic transition and statistical mechanical
randomness, while in classical system, the randomness only comes
from the later case. We use the two-level system as an illustration
to demonstrate the validity of quantum Crooks FT and quantum JE
beyond the linear response regime.

Before concluding the paper, we would like to mention the following
points. First, though the quantum non-adiabatic transition is
introduced into the controlling substep, this substep is time
reversal symmetric. I. e., all the time asymmetry is due the
relaxation substep (statistical mechanical randomness), rather than
the controlling substep (quantum non-adiabatic transition). This is
the same as the classical case. Second, when we change the
Hamiltonian slowly, we reproduce the proof of Crooks for classical
systems. In this sense, we say that our proof includes the classical
Crooks FT and classical JE as a special case. Third, for classical
system, the Crooks FT and JE have been experimentally verified
\cite{classicalexperiment}. However, for a quantum mechanical
system, the experimental exploration on Crooks FT and JE has not
been reported (an exception is \cite{schmidt}). This perhaps is
mainly due to the fact that microscopic work in a quantum mechanical
system is not a well defined observable \cite{hanggi}. There is no
well defined pressure or force for a quantum system \cite{quan0811}.
Hence, we cannot follow the way that we do in classical system to
measure the force and make the integral of the force by the
extension. On the contrary, we will have to introduce quantum
measurement processes to confirm the initial and final energy of the
system and then calculate the microscopic work done from the
difference of the initial and final energy difference
\cite{mukemal}. Fourth, though the numerical simulations consider
only the special cases: 1) the system reach thermal equilibrium with
the heat bath in every relaxation substep, and 2) the quantum
adiabatic conditions are satisfied in every controlling substep, the
quantum Crooks FT and quantum JE are not constrained in these
special cases. Finally, our numerical simulations based on a
two-level system can possibly be testified by employing Josephson junction
charge qubit \cite{chargequbit}. Discussion about employing
Josephson Junction qubit to test the quantum Crooks FT and quantum
JE will be given later.

\section{acknowledgments}

HTQ thanks Wojciech H. Zurek, G. Crooks and Rishi Sharma for stimulating discussions and gratefully acknowledges the support of the U.S. Department of Energy through
the LANL/LDRD Program for this work.


\begin{thebibliography}{99}

\bibitem{nonequilibrium} S. R. de Groot and P. Mazur, \emph{Nonequilibrium Thermodynamics}, (North-Holland, Amsterdam, 1962).

\bibitem{evans}  D. J. Evans and D. J. Searles, Phys. Rev. E \textbf{50}, 1645 (1994); D. J. Evans and D. J. Searles, Advances in Physics, \textbf{51}, 1529 (2002).

\bibitem{JE}  C. Jarzynski, Phys. Rev. Lett. \textbf{78}, 2690 (1997).

\bibitem{crooks}  Crooks, J. Stat. Phys. \textbf{90}, 1481 (1998); G. E. Crooks, Phys. Rev. E \textbf{60}, 2721 (1999); Gavin E. Crooks, Phys. Rev. E \textbf{61}, 2361 (2000).

\bibitem{followup} C. Bustamante, J. Liphardt, and F. Ritort, Phys. Today, \textbf{54}, (7) 43 (2005); M. Haw, Phys. World, \textbf{20}, (11) 25, (2007); C. Jarzynski, Eur. Phys. J. B.  \textbf{64}, 331 (2008) and reference therein.

\bibitem{classicalnumerical} D. J. Evans, E.G.D. Cohen, and G.P. Morriss, Phys. Rev. Lett. \textbf{71}, 2401 (1993); C. Jarzynski, Phys. Rev. E \textbf{56}, 5018 (1997).

\bibitem{classicalexperiment} G. M. Wang, E. M. Sevick, E. Mittag, D. J. Searles, and D. J. Evans, Phys. Rev. Lett. \textbf{89}, 050601 (2002); D. M. Carberry, J. C. Reid, G. M. Wang, E. M. Sevick, D. J. Searles, and Denis J. Evans, Phys. Rev. Lett. \textbf{92}, 140601 (2004); J. Liphardt, S. Dumont, S.B. Smith, I. Tinoco Jr., C. Bustamante, Science, \textbf{296}, 1832 (2002); D. Collin, F. Ritort, C. Jarzynski, S.B. Smith, I. Tinoco Jr., C. Bustamante, Nature  \textbf{437}, 231 (2005); N. C. Harris, Y. Song, Ching-Hwa Kiang, Phys. Rev. Lett.  \textbf{99}, 068101 (2007).

\bibitem{quantumJE} S. Yukawa, J. Phys. Soc. Jpn \textbf{69}, 2367 (2000); J. Kurchan, arXiv:cond-mat/0007360v2; H. Tasaki, arXiv:cond-mat/0009244v2; V. Chernyak, S. Mukamel, Phys. Rev. Lett. \textbf{93}, 048302 (2004); M. Esposito, and S. Mukamel, Phys. Rev. E. \textbf{73}, 046129 (2006); P. Talkner, P. H\"{a}nggi, M. Morillo, arXiv:0707.2307v1; J. Teifel, G. Mahler, Phys. Rev. E \textbf{76}, 051126 (2007); H. Schroder, J. Teifel, G. Mahler, Eur. Phys. J. Special Topics, \textbf{151}, 181 (2007); P. Talkner, M. Campisi, and P. H\"{a}nggi, arXiv:0811.0973v1;

\bibitem{isolatedcrooks} P. Talkner, P. H\"{a}nggi, J. Phys. A.: Math. Theor. 40, F569 (2007); S. Deffner, and E. Lutz, Phys. Rev. E \textbf{77}, 021128 (2008); P. Talkner, P. H\"{a}nggi, and M. Morillo, Phys. Rev. E \textbf{77}, 051131 (2008).

\bibitem{schmidt}  G. Huber, F. Schmidt-Kaler, S. Deffner, E. Lutz, Phys. Rev. Lett. \textbf{101}, 070403 (2008).

\bibitem{reverse} For classical systems, if the forward process is described by a trajectory in the phase space $(\vec{p}_{0}, \vec{q}_{0}) \rightarrow (\vec{p}_{1}, \vec{q}_{1}) $ as the Hamiltonian is changed from $H(\lambda_{0})$ to $H(\lambda_{1})$. The time-reversed trajectory is $(-\vec{p}_{1}, \vec{q}_{1}) \rightarrow (-\vec{p}_{0}, \vec{q}_{0}) $ as the Hamiltonian is changed from $H(\lambda_{1})$ to $H(\lambda_{0})$. For quantum systems, if the forward trajectory is $\left \vert  \psi(t_{0}) \right\rangle \rightarrow \left \vert  \psi(t_{1}) \right\rangle$ as the Hamiltonian is changed from $H(\lambda_{0})$ to $H(\lambda_{1})$, the time-reversed trajectory is $\Theta \left \vert  \psi(t_{1}) \right\rangle \rightarrow \Theta \left \vert  \psi(t_{0}) \right\rangle$ when the Hamiltonian is changed from $H(\lambda_{1})$ to $H(\lambda_{0})$ \cite{sakurai}.

\bibitem{sakurai} J. J. Sakurai, \emph{Modern Quantum Mechanics} (Revised Edition), (Reading, Addison-Wesley, 1994).

\bibitem{chandler} D. Chandler,  \emph{Introduction to Modern Statistical Mechanics}, (Oxford University Press, New York, 1987).

\bibitem{theta} C. Jarzynski, and D. K. Wojcik, Phys. Rev. Lett. \textbf{92}, 230602 (2004); W. De Roeck, C. Maes, Phys. Rev. E 69, 026115 (2004);  T. Monnai, Phys. Rev. E \textbf{72}, 027102 (2005); G. E. Crooks, Phys. Rev. A 77, 034101 (2008); D. Andrieux and P. Gaspard, Phys. Rev. Lett. \textbf{100}, 230404 (2008).

\bibitem{quan08} H. T. Quan. S. Yang, and C. P. Sun, Phys. Rev. E. \textbf{78}, 021116 (2008).

\bibitem{mukemal}  S. Mukamel, Phys. Rev. Lett. \textbf{90}, 170604 (2003).

\bibitem{quan0811}  H. T. Quan, arXiv: 0811.2756.

\bibitem{hanggi} P. Talkner, E. Lutz, and P. H\"{a}nggi, Phys. Rev. E \textbf{75}, 050102(R) (2007).

\bibitem{chargequbit} J. Q. You, and F. Nori, Phys. Today \textbf{58}, No. 11, 42 (2005).

\end{thebibliography}
\end{document}